\documentclass[amsmath,amssymb,showpacs,floatfix, aps, prl, twocolumn, superscriptaddress]{revtex4-1}
\usepackage{graphicx}
\usepackage{dcolumn}
\usepackage{bm}
\usepackage{color}
\usepackage{hyperref}
\hypersetup{colorlinks=true,citecolor=blue,linkcolor=blue, urlcolor=blue}

\begin{document}
\title{Ab initio Electron Mobility and Polar Phonon Scattering in GaAs}

\author{Jin-Jian Zhou}
\affiliation {Department of Applied Physics and Materials Science, Steele Laboratory, 
California Institute of Technology, Pasadena, California 91125, USA}

\author{Marco Bernardi}
\email{bmarco@caltech.edu}
\affiliation {Department of Applied Physics and Materials Science, Steele Laboratory, 
California Institute of Technology, Pasadena, California 91125, USA}
\date{\today}

\begin{abstract}
\noindent
In polar semiconductors and oxides, the long-range nature of the electron-phonon (\textit{e}-ph) interaction is a bottleneck to compute charge transport from first principles. 
Here, we develop an efficient ab initio scheme to compute and converge the \textit{e}-ph relaxation times (RTs) and electron mobility in polar materials. 
We apply our approach to GaAs, where using the Boltzmann equation with state-dependent RTs, we compute mobilities in excellent agreement with experiment at 250$-$500~K. 
The $e$-ph RTs and the phonon contributions to intravalley and intervalley $e$-ph scattering are also analyzed.
Our work enables efficient ab initio computations of transport and carrier dynamics in polar materials.
\end{abstract}
\maketitle
 
Semiconductors with polar bonds, such as III-V and II-VI compounds, and oxides, are important in condensed matter physics and for technological applications.
Charge transport in these polar materials plays a key role in electronics, optoelectronic, photovoltaics, and photocatalysis. 
%
Novel experiments~\cite{Verma2014,Yu2016} are dramatically advancing understanding of charge transport in polar materials. 
Yet, their microscopic interpretation requires detailed knowledge of the carrier scattering processes. 
Since fabricating pure crystals is challenging for many polar materials, 
extracting intrinsic charge transport properties from experiment is non-trivial; questions related to the role of doping, impurities, 
and defects often arise when interpreting transport measurements.\\
\indent
Ab initio computational approaches to study carrier transport and scattering mechanisms are uniquely poised to advance understanding of polar materials. 
Ab initio calculations of carrier mobility \cite{Restrepo2009,Kaasbjerg2012,Li2013,Park2014,Li2015,Mustafa2016,Gunst2016} and scattering \cite{Bernardi2014,Bernardi2015,Bernardi2015a,Tanimura2016} are a recent development, 
and have been reported so far only for a few metals and nonpolar semiconductors. 
%
%
\mbox{However,} the mobility in polar semiconductors and oxides, which is the focus of this Rapid Communication,  is still typically investigated with semi-empirical models~\cite{Verma2014,Faghaninia2015}.
One major challenge in computing transport in polar materials is the Fr\"ohlich interaction \cite{Froehlich1954}, a long-range coupling between electrons and longitudinal optical (LO) phonons. 
Electron-phonon (\textit{e}-ph) scattering due to LO modes, and in general to polar phonons (PPs), is typically the main carrier scattering mechanism in polar materials, 
but it cannot yet be included in ab initio transport calculations.\\
\indent
%
%
%
%
When computed directly using density functional perturbation theory (DFPT)~\cite{Baroni2001}, the \textit{e}-ph matrix elements correctly include 
the Fr\"ohlich interaction for arbitrary values of the phonon wavevector $\mathbf{q}$. 
Yet, the very large number of $e$-ph matrix elements necessary to converge the mobility and the $e$-ph relaxation times (RTs)~\cite{Bernardi2014,Bernardi2015,Bernardi2015a} 
make direct DFPT calculations impractical due to computational cost. 
%
%
For metals and nonpolar semiconductors, in which the \textit{e}-ph interaction is short-ranged, Wannier interpolation 
\cite{Giustino2007} can be employed to obtain \textit{e}-ph matrix elements on fine Brillouin zone (BZ) grids. 
In polar materials, Wannier interpolation is inconvenient since the $e$-ph matrix elements for LO modes diverge \mbox{as $1/q$ for $\mathbf{q}\rightarrow$ 0}. 
%
A method was recently proposed \cite{Sjakste2015,Verdi2015} to split the \textit{e}-ph matrix elements $g$ into short- and long-range parts, $g = g^{S} \!+ g^{L}$. The long range part $g^{L}$ containing the $1/q$ singularity is the ab initio generalization of the $e$-ph Fr\"ohlich interaction \cite{Verdi2015}, 
and can be evaluated using an analytical formula based on the Vogl model \cite{Vogl1976}. 
The short-range part $g^{S}$ is well-behaved, and can be computed by Wannier interpolation. 
This method can correctly reproduce the $e$-ph matrix elements computed with DFPT for arbitrary values of $\mathbf{q}$ \cite{Sjakste2015,Verdi2015}. 
It has also been used to compute the $e$-ph RTs for specific electronic states \cite{Sjakste2015,Verdi2015} for showcasing the computations possible with this important approach. 
\mbox{However,} ab initio computations of charge transport, which involve the daunting task of computing and converging the $e$-ph RTs on fine grids in the entire BZ, have yet to be carried out in polar bulk materials.
Computations of \textit{e}-ph scattering have recently appeared for polar two-dimensional materials~\cite{Kaasbjerg2012,Gunst2016,Sohier2016}, where since the Fr\"ohlich interaction is well-behaved at small $\mathbf{q}$, the computational challenges are similar to those of nonpolar bulk materials.\\
\indent
%
In this Rapid Communication, we present fully ab initio calculations of electron mobility in a polar bulk material. 
An efficient scheme to compute and converge the \textit{e}-ph RTs on fine BZ grids is derived. 
We apply this approach to GaAs, a polar material for which accurate mobility measurements are available. 
The Boltzmann equation within the RT approximation is employed, in combination with temperature- and state-dependent RTs, to compute the electron mobility for temperatures of 200$-$700 K, 
achieving excellent agreement with experiment (e.g., within 5\% of experiment at 300 K). 
We analyze the phonon mode contributions to the RTs and mobility, and find that PPs dominate intravalley scattering and transport, 
while acoustic phonons dominate intervalley scattering and hot carrier dynamics. 
%
%
%
Our work enables ab initio transport calculations in polar materials at roughly the same computational cost as in nonpolar materials, 
and advances the microscopic understanding of carrier dynamics in GaAs.\\
\indent
%
%
We carry out density functional theory (DFT) calculations on GaAs with a relaxed lattice constant of 5.55 \AA, using the local density approximation (LDA) \cite{Perdew1981} and a plane wave basis with the {\sc Quantum ESPRESSO} code \cite{Giannozzi2009}. Norm-conserving pseudopotentials \cite{Troullier1991} and a plane-wave kinetic energy cutoff of 72 Ry are employed to obtain the ground state charge density and bandstructure.  
We use DFPT to compute the lattice dynamical properties \cite{Baroni2001} and the \textit{e}-ph matrix elements, $g_{nm\nu}(\mathbf{k},\mathbf{q})$, on coarse 8$\times$8$\times$8 $\mathbf{k}$- and $\mathbf{q}$-point BZ grids. 
These $e$-ph matrix elements represent the transition amplitudes from a Bloch state with band index $n$ and crystal momentum $\mathbf{k}$ to a Bloch state with quantum numbers $m$ and $\mathbf{k}+\mathbf{q}$, mediated by the emission or absorption of a phonon with branch index $\nu$ and wavevector $\mathbf{q}$ \cite{Bernardi2016a}. 
The $e$-ph matrix elements for arbitrary $\mathbf{k}$- and $\mathbf{q}$-points are then obtained by adding the short-range part $g^S_{nm\nu}(\mathbf{k},\mathbf{q})$, 
obtained by Wannier interpolation, and the long-range part $g^L_{nm\nu}(\mathbf{k},\mathbf{q})$, 
which we independently implement using the method in Ref.~\cite{Verdi2015}.
%
%
The band- and $\mathbf{k}$-dependent \textit{e}-ph scattering rate $\Gamma_{n\mathbf{k}}^{\mathrm{e-ph}}$ is computed with an in-house modified version of the EPW code \cite{Noffsinger2010}, from the imaginary part of the lowest order \textit{e}-ph self-energy, $\mathrm{Im}\Sigma_{n\mathbf{k}}^{\mathrm{e-ph}}$, 
using $\Gamma_{n\mathbf{k}}^{\mathrm{e-ph}} = 2/\hbar \mathrm{Im}\Sigma_{n\mathbf{k}}^{\mathrm{e-ph}}$ \cite{Bernardi2016a}: 
\begin{widetext}
\begin{equation}
\Gamma_{n\mathbf{k}}^{\mathrm{e-ph}} (T) = 
\frac{2\pi}{\hbar} \sum_{m \nu\mathbf{q}}\left |g_{nm\nu}(\mathbf{k},\mathbf{q}) \right|^{2} 
[\left(N_{\nu \mathbf{q}} + 1 - f_{m \mathbf{k}+\mathbf{q}}\right)\delta\!\left(\varepsilon_{n\mathbf{k}}-\varepsilon_{m\mathbf{k}+\mathbf{q}}-\hbar\omega_{\nu\mathbf{q}}\right) 
\,+\, \left(N_{\nu \mathbf{q}} + f_{m \mathbf{k}+\mathbf{q}}\right)\delta\!\left(\varepsilon_{n\mathbf{k}}-\varepsilon_{m\mathbf{k}+\mathbf{q}}+\hbar\omega_{\nu\mathbf{q}}\right)]
\label{eq:selfenergy}
\end{equation}
\end{widetext}
where $T$ is the temperature, $\varepsilon_{n\mathbf{k}}$ and $\hbar\omega_{\nu\mathbf{q}}$ the electron and phonon energies, respectively, and $f_{n\mathbf{k}}$ and $N_{\nu\mathbf{q}}$ the corresponding occupations. 
Here, the temperature dependence is included in the occupations, while the $e$-ph matrix elements are computed in the ground state. The $e$-ph RTs, $\tau_{n\mathbf{k}} \!=\! (\Gamma_{n\mathbf{k}}^{\mathrm{e-ph}})^{-1}$, are the inverse of the scattering rates. \\
%
%
\indent
The electrical conductivity $\sigma$ is computed within the RT approximation of the Boltzmann transport equation \cite{Mahan2010,Pizzi2014},
\begin{equation}
\sigma_{\alpha\beta} = e^{2} \int_{-\infty}^{+\infty}{dE}(-\partial{f} / \partial{E})\Sigma_{\alpha\beta}(E,T) \,.
\label{eq:cond}
\end{equation}
%
%
$\Sigma_{\alpha\beta}(E,T)$ is the transport distribution function (TDF) at energy $E$ and temperature $T$,
\begin{equation}
\Sigma_{\alpha\beta}(E,T) = \frac{2}{V_{\text{uc}}} \sum_{n\mathbf{k}}{ \tau_{n\mathbf{k}}(T)\mathbf{v}_{n\mathbf{k}}^{\alpha} \mathbf{v}_{n\mathbf{k}}^{\beta}\delta(E-\varepsilon_{n\mathbf{k}}) } ,
\label{eq:tdf}
\end{equation}
computed here with a tetrahedron integration method \cite{Bloechl1994}, using ab initio \textit{e}-ph RTs and interpolated \cite{Yates2007,Mostofi2014} band velocities $\mathbf{v}_{n\mathbf{k}}$; $V_{\text{uc}}$ is the unit cell volume.
The mobility is obtained as $\mu = \sigma / n_{C}e$, where $n_{C}$ is the intrinsic carrier concentration.\\
\indent
%
%
%
\begin{figure}
\includegraphics[width=\columnwidth]{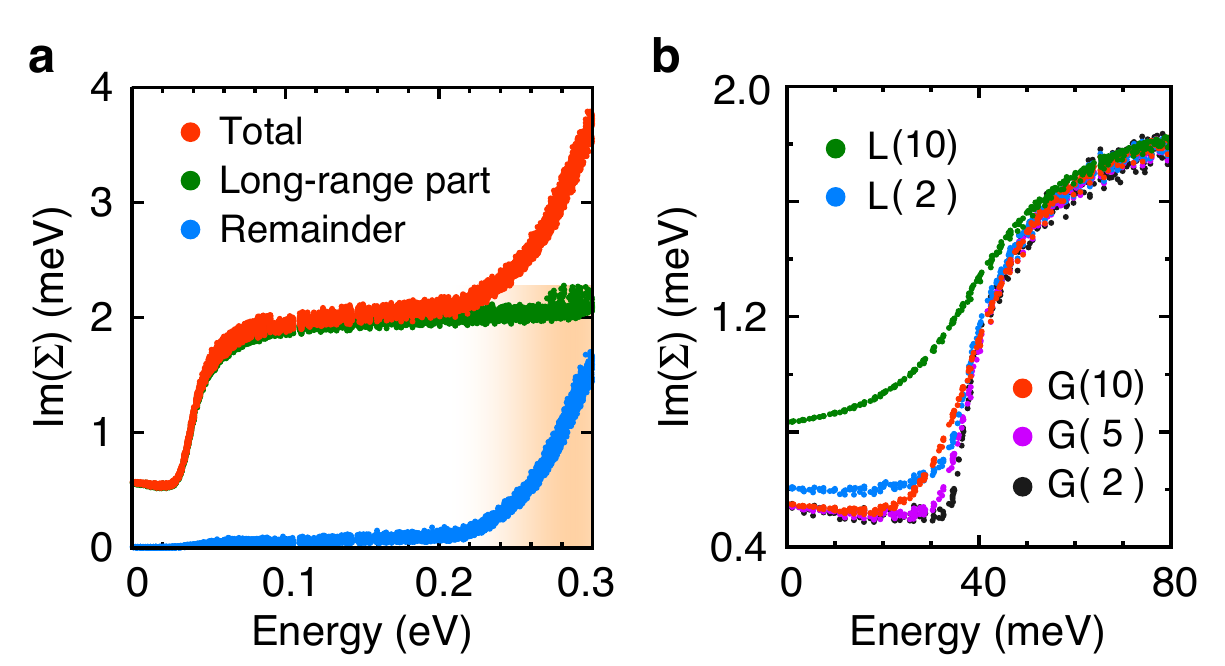}
\caption{(Color online) (a) The converged scattering rate in Eq.~(\ref{eq:selfenergy}) (curve labeled as \lq\lq Total\rq\rq), expressed in terms of $\mathrm{Im}\Sigma^{\text{e-ph}}$ in meV units. 
Shown are also the long-range and remainder contributions, which add up to the total. 
(b) Convergence of the scattering rate near the CBM. Shown are the results for Lorentzian (L) and Gaussian (G) broadenings. The broadening parameter, in meV units, is given in parentheses.  
}\label{fig1}
\end{figure}
We first discuss our approach for efficiently computing the $e$-ph scattering rates in Eq.~(\ref{eq:selfenergy}) in polar materials. 
%
%
%
%
Due to the $1/q$ singularity of the long-range part $g^L$ (dropping all the indices from now on),   
converging $\Gamma_{n\mathbf{k}}^{\mathrm{e-ph}}$ when using $g = g^S + g^L$ is computationally prohibitive 
since the sum over $\mathbf{q}$ in Eq.~(\ref{eq:selfenergy})converges very slowly. 
We reason that the matrix elements $g^L$ are inexpensive to compute, as they merely require evaluating an analytical function at one point \cite{Verdi2015}. 
While converging  $\Gamma_{n\mathbf{k}}^{\mathrm{e-ph}}$ using $g\!=\!g^L$ alone requires as many as $10^6$$-$10$^7$ $\mathbf{q}$ points due to the singularity, this task is still relatively inexpensive. 
On the other hand, computing each short-range $e$-ph matrix element $g^S$ is rather costly as it requires Wannier interpolation. 
For example, converging $\Gamma_{n\mathbf{k}}^{\mathrm{e-ph}}$ using $g\!=\!g^S$ alone, as done for nonpolar materials \cite{Bernardi2014,Bernardi2015,Bernardi2015a}, 
typically requires $\sim$10$^3$$-$10$^5$ $\mathbf{q}$ points and is computationally very expensive.\\
\indent
%
%
On this basis, we split $|g|^{2}$ in Eq.~(\ref{eq:selfenergy}) into two parts, the long-range part $|g^L|^{2}$ and the remainder $( |g|^{2} - |g^L|^{2} )$. 
Equation~(\ref{eq:selfenergy}) with $|g|^2\!=\!|g^L|^{2}$ and $|g|^2\!=\!( |g|^{2} - |g^L|^{2} )$ is then used to 
separately compute the long-range and remainder contributions to $\Gamma_{n\mathbf{k}}^{\mathrm{e-ph}}$, respectively, which add up to the total scattering rate. 
%
Figure~\ref{fig1}(a) shows the long-range, remainder, and total scattering rates in GaAs at 300 K.
Each contribution is computed and converged separately, with important advantages for the choice of the integration grids. 
For the long-range part, we treat the $1/q^2$ singularity of $|g^L|^2$ by using Monte Carlo integration with importance sampling near the BZ center, using $\mathbf{q}$ points randomly sampled from a Cauchy distribution~\cite{[{We sample each component of \textbf{q} within [-0.5, 0.5] using the Cauchy distribution, with a probability density function $P(x)=\frac{1}{\pi}\frac{\varepsilon}{x^{2}+\varepsilon^{2}}$. We employ $\varepsilon=0.035$ in our calculations.}] cauchy_dist}.
For the remainder part, convergence requires $\sim$10$^3$$-$10$^5$ $\mathbf{q}$ points as in nonpolar materials, 
and is achieved incrementally using Monte Carlo integration over multiple random $\mathbf{q}$-point grids \cite{Bernardi2015,Bernardi2015a}.  \\
\indent
\mbox{Overall,} the approach enables calculations of $e$-ph RTs in polar materials at roughly the same cost as in nonpolar materials, with a small overhead to compute the long-range contribution. 
For comparison, converging $\Gamma_{n\mathbf{k}}^{\mathrm{e-ph}}$ in Eq.~(\ref{eq:selfenergy}) directly with $g \!=\! g^S \!+ g^L$ is dramatically more expensive, by a factor equal to the ratio $N_L / N_R \approx 10$$-$1,000 between the number of $\mathbf{q}$ points 
needed to converge the long-range ($N_L\approx 10^6$$-$$10^7$) and the remainder ($N_R \approx 10^3$$-$$10^5$) parts. 
Our idea of dividing and conquering the long-range part thus enables fast computations of the $e$-ph RTs in polar materials.\\
\indent
%
%
The approximation employed for the $\delta$ function in Eq.~(\ref{eq:selfenergy}) is also crucial to converge the scattering rate, especially near the conduction band minimum (CBM). 
We use \mbox{$\delta(x) = \lim_{\eta \to 0} f(x,\eta)$}, where $\eta$ is a small broadening parameter, and test both Lorentzian and Gaussian broadenings, 
with distributions $f(x,\eta) = \frac{1}{\pi}\frac{\eta}{x^{2}+\eta^{2}}$ and $f(x,\eta) = \frac{1}{\sqrt{\pi}}\frac{1}{\eta}e^{-(\frac{x}{\eta})^{2}}$, respectively.
Convergence of $\Gamma_{n\mathbf{k}}^{\mathrm{e-ph}}$ in Eq.~(\ref{eq:selfenergy}) is achieved by choosing a small value of $\eta$ (e.g., 10 meV) and using a number of $\mathbf{q}$ points $N_{\mathbf{q}}(\eta)$  
large enough to converge the sum over $\mathbf{q}$ for the given value of $\eta$. 
Existence of the limit guarantees that upon decreasing $\eta$ to a new value, and increasing $N_{\mathbf{q}}(\eta)$ accordingly, the scattering rate no longer varies as $\eta$ is decreased further.  
We employ both Lorentzian and Gaussian broadenings, with parameters $\eta$ of 2, 5, and 10~meV, and for each case we  converge the scattering rate with respect to the number of $\mathbf{q}$ points.\\
\indent
%
%
%
The results of this convergence study are shown in Fig.~\ref{fig1}(b) for energies up to $\sim$0.1 eV above the CBM (from now on, we reference the electron energy to the CBM). 
We find that the scattering rate for low energy electrons in the $\Gamma$ valley is highly sensitive to the broadening. 
In particular, Lorentzian broadening tends to overestimate the scattering rate even for a small value of $\eta=2$~meV.  
Gaussian broadening is easier to converge: A small parameter $\eta\approx$ 5~meV is sufficient to converge the scattering rate in the $\Gamma$ and $L$ valleys. 
As shown below, electronic states in this energy range play a crucial role in transport. 
Note that even a relatively small 10 meV Lorentzian broadening, as typically employed, would lead to enormous errors in the mobility. 
On the other hand, a 10 meV Lorentzian broadening is acceptable at energy above $\sim$0.3 eV, as electronic states with higher energy are less sensitive to broadening. 
In what follows, we employ a 5~meV Gaussian broadening.\\
\indent
%
%
\begin{figure}
\includegraphics[width=\columnwidth]{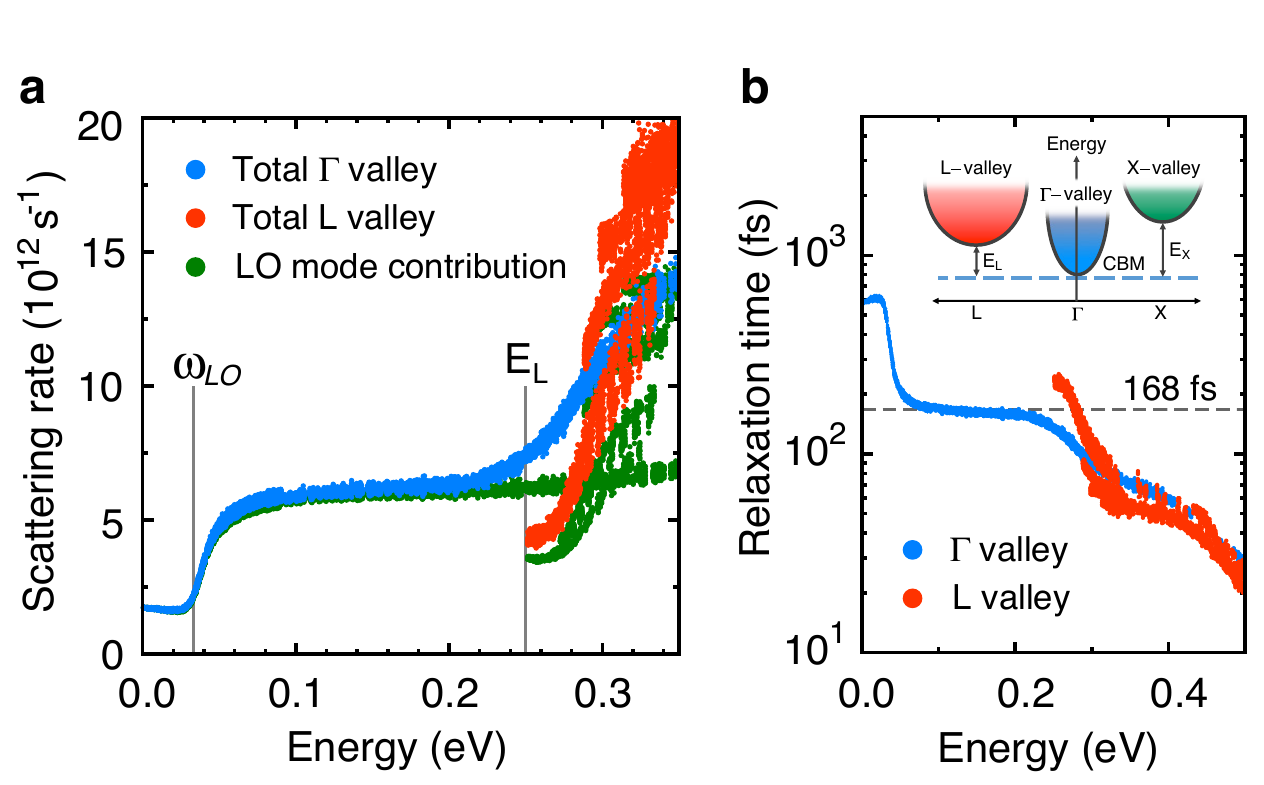}
\caption{(Color online) (a) \textit{e}-ph scattering rate and (b) RT for electrons in GaAs with energies within $\sim$0.4 eV of the CBM, which is the origin of the energy axis. Data points in blue (red) are for electronic states in the $\Gamma$ ($L$) valley. 
The scattering rate associated with LO phonon scattering alone is also shown in (a). The inset in (b) shows a schematic of the $\Gamma$ valley, and of the $L$ and $X$ valleys in GaAs, with energies of $E_{L}\!\approx\!0.25$ eV and $E_{X}\!\approx\!0.45$ eV above the CBM, respectively.
}\label{fig2}
\end{figure}
%
%
The conduction band of GaAs has a multi-valley character, as sketched in the inset of Fig.~\ref{fig2}(b). The minima of the $L$ and $X$ valleys are at energies $E_{L}\!\approx\!0.25$ eV
and $E_{X}\approx0.45$ eV above the CBM at $\Gamma$, respectively~\cite{[{See Supplemental Material for computed band structure of GaAs}] supp_mat}.
We first focus on $e$-ph scattering in the $\Gamma$ and $L$ valleys, which is of crucial importance to compute charge transport in GaAs. 
Figure~\ref{fig2}(a) shows the \textit{e}-ph scattering rate at 300~K within $\sim$0.4 eV of the CBM, separately for electronic states in the $\Gamma$ and $L$ valleys.
%
For electrons in the $\Gamma$ valley, only intravalley scattering is possible for energies up to $E_{L}$.
Small-$\mathbf{q}$ LO phonon scattering dominates in this energy range, 
as shown in Fig.~\ref{fig2}(a) by comparing the total $e$-ph scattering rate with the one due to LO phonons alone. 
The scattering rate is nearly constant over the 0.05$-$0.25 eV energy range, with an associated RT [see Fig.~\ref{fig2}(b)] of $\sim$168 fs. 
Our RT at 300 K is excellent agreement with room temperature experiments, e.g., $\sim$165 fs in Ref. \cite{Kash1985}.
%
%
%
%
%
%
At energies below $\sim$0.05 eV the scattering rate drops sharply, and approaches the CBM with a constant trend. 
Within $\hbar\omega_{LO} \approx 35$ meV of the CBM, the phase space for LO phonon emission vanishes, and the scattering process is dominated by LO phonon absorption. 
The scattering rate in this energy range is roughly proportional to the LO phonon occupation, and is strongly temperature dependent. 
Our computed RT for LO phonon absorption at 300~K is $\sim$600 fs [Fig.~\ref{fig2}(b)].
%
%
%
\begin{figure}
\includegraphics[scale=0.8]{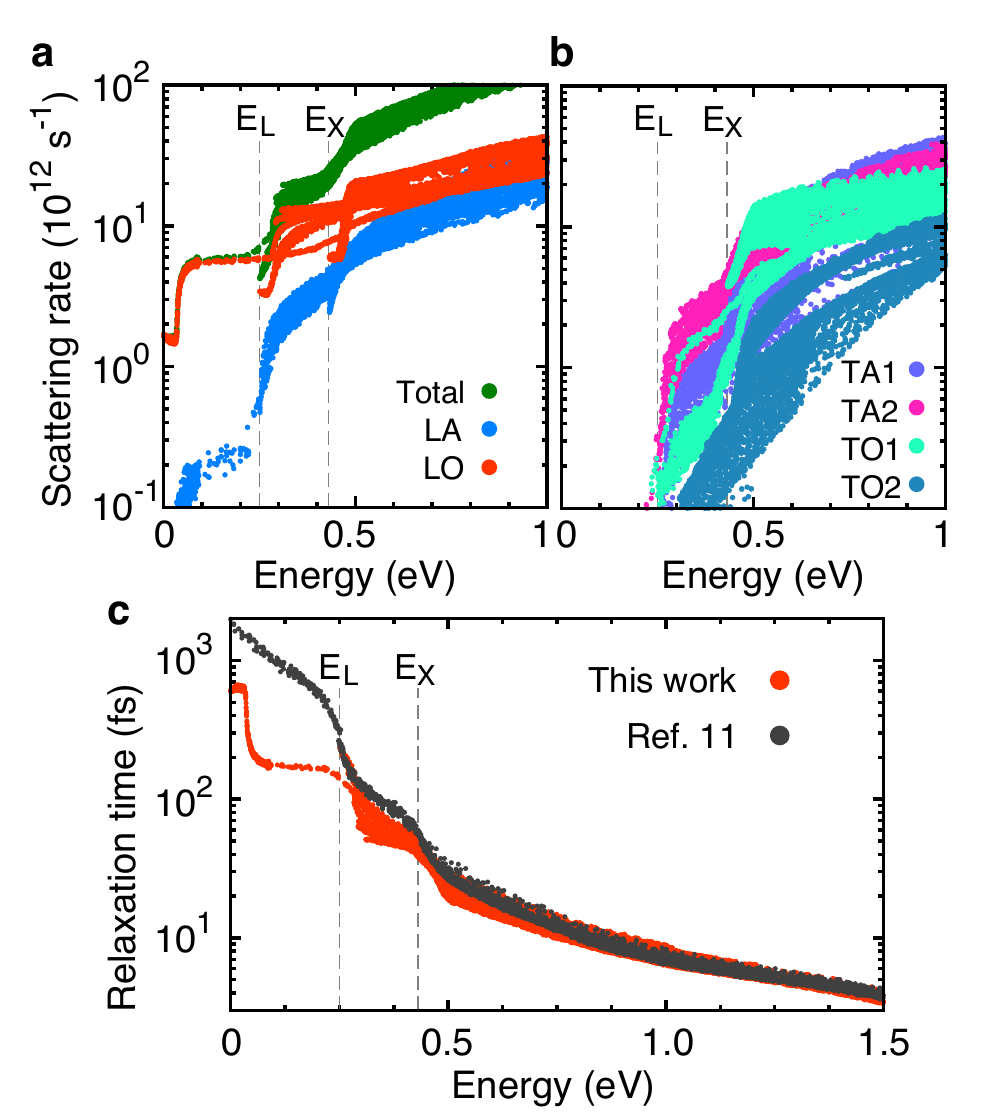}
\caption{(Color online) Mode resolved \textit{e}-ph scattering rates. (a) Total scattering rate, shown along with the contributions from the LO and LA modes alone. 
(b) Contributions from the transverse modes. 
(c) Comparison of ab initio RTs including the PP scattering (red curve), as computed in this work, with previous calculations 
in Ref.~\cite{Bernardi2015} that neglect PP scattering (gray curve). 
}\label{fig3}
\end{figure}
%
%
%
At energy higher than $E_{L}$, $\Gamma$$-$$L$ intervalley scattering becomes possible, and the scattering rate increases rapidly as a result. 
Intravalley scattering in the $L$ valley, also possible above $E_{L}$, is dominated by PP scattering.   
It exhibits a scattering rate with multiple branches [Fig.~\ref{fig2}(a)], and thus a strong $\mathbf{k}$-dependence,  
due to the anisotropy of the $L$ valley.\\
\indent
%
%
To gain additional insight into $e$-ph scattering, we plot in Figs.~\ref{fig3}(a) and \ref{fig3}(b) the contributions from different phonon modes to the total scattering rates, over a wider energy 
range (up to $\sim$1 eV) than analyzed above. 
Though LO scattering is dominant in the $\Gamma$ and $L$ valleys (below and above $E_L$, respectively), the longitudinal acoustic (LA) mode also contributes to small-$\mathbf{q}$ intravalley scattering through the so-called piezoelectric interaction~\cite{Yu2010,[{Note that we neglect the quadrupole term proportional to the second derivative of the effective charges with respect to the phonon wavevector (see eq. 3.15 in Ref.~\cite{Vogl1976}). However, as shown in Fig. S2(b) of the Supplemental Material~\cite{supp_mat}, comparison with our DFPT calculation shows that this term is negligibly small in GaAs.}] piezo_eph}.  
For energies between $E_{L}$ and $E_{X}$, the $\Gamma$$-$$L$ intervalley scattering is dominated by large-$\mathbf{q}$ LA and transverse acoustic (TA) phonon scattering. 
\begin{figure}
\includegraphics[scale=0.8]{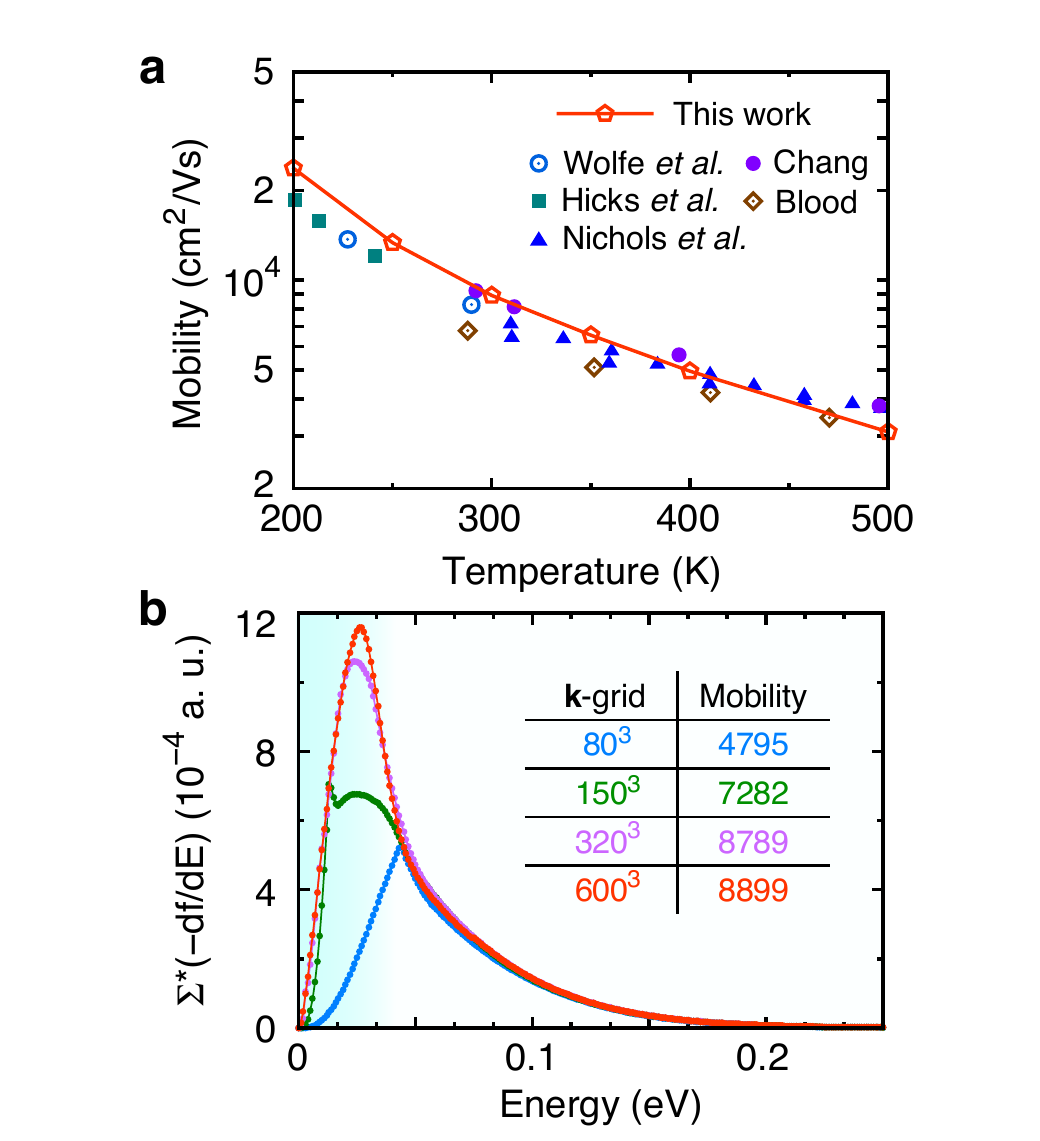}
\caption{(Color online) (a) Electron mobilities close to room temperature. The experimental values are taken from Refs.~\cite{Rode1970,*Rode1971, Blakemore1982, *Hicks1969, *Wolfe1970, *Blood1972, *Nichols1980}, and compared with our computed ab initio mobilities (red line). (b) Convergence of the integrand in Eq.~(\ref{eq:cond}) and the mobility (values given as a table) at 300~K with respect to the fine $\mathbf{k}$ grids used for the tetrahedron integration in Eq.~(\ref{eq:tdf}).
The blue, green, purple and red curves are computed using fine grids of 80$^3$, 150$^3$, 320$^3$, 600$^3$ $\mathbf{k}$ points in the BZ, respectively. 
}\label{fig4}
\end{figure}
At energy greater than $E_{X}$, the TA modes are the main source of $e$-ph scattering, consistent with recent results \cite{Bernardi2015}.\\
\indent
Figure~\ref{fig3}(c) compares our computed $e$-ph RTs with those obtained in previous work \cite{Bernardi2015} that did not include PP scattering as it focused on hot carriers with high energy above the CBM. 
We note that hot carrier calculations in GaAs have also appeared in Ref.~\cite{Tanimura2016}, which, similar to Ref.~\cite{Bernardi2015}, focused on higher carrier energies than those of interest here.
For energies above $E_{X}$, we find that the change in the RTs due to PP scattering is rather small, consistent with the fact that large-$\mathbf{q}$ scattering dominates in this energy range. 
PP scattering is thus almost negligible in hot carrier dynamics, and the conclusion that carriers excited above $E_X$ in GaAs thermalize 
chiefly by emitting acoustic phonons \cite{Bernardi2015} is still valid when PP scattering is included.
%
%
%
However, Fig.~\ref{fig3}(c) also shows that for electronic states with energy lower than $E_{L}$, the inclusion of PP scattering makes a dramatic difference in the RTs. 
PP scattering additionally leads to a strong $\mathbf{k}$-dependence of the RTs for energies between $E_{L}$ and $E_{X}$. 
These effects are crucial to accurately compute electron mobility and transport.\\
\indent
Next, we discuss the phonon-limited mobility in GaAs, as shown in Fig.~\ref{fig4}(a). For temperatures between 200 and 500~K, 
%
%
%
our computed mobilities are in excellent agreement with experiment; for example, our room-temperature result is $\sim$8900 cm$^{2}$/Vs, versus experimental values of 8200$-$8900 cm$^{2}$/Vs~\cite{Hicks1969,Rode1971}.
Converging the electron mobility is very challenging since Eq.~(\ref{eq:tdf}) requires BZ integration on very fine $\mathbf{k}$-point grids. We employ the tetrahedron integration method to converge the conductivity with high accuracy.
To investigate the convergence of our mobility calculations, we plot the integrand in Eq.~(\ref{eq:cond}), $(-\partial{f}/\partial{E})\Sigma_{\alpha\alpha}(E, T)$, at $T= 300$~K;  
this function is proportional to the TDF, and is employed to visualize the contributions to the conductivity from electronic states at different energies.  
The integrands calculated using four different choices of the $\mathbf{k}$-point grids, together with the corresponding mobilities, are shown in Fig.~\ref{fig4}(b). 
We find that the main contribution to the mobility originates from electronic states in a small energy window (at room temperature, $\sim$0.05 eV) above the CBM, 
where scattering is dominated by LO phonon absorption. 
Extremely fine grids are necessary to sample this small BZ region and capture the rapid changes of the RTs near the CBM. 
Figure~\ref{fig4}(b) shows that convergence of the mobilities is achieved only for grids with more than 600$^3$ $\mathbf{k}$ points, and that even fine grids with 150$^3$ $\mathbf{k}$ points lead to large errors in the mobility calculation.
Previous theoretical work \cite{Rode1970,*Rode1971} using empirical models concluded that iterative methods beyond the RT approximation 
are necessary to obtain mobilities in agreement with experiment. However, 
our results demonstrate that if ab initio temperature- and state-dependent RTs are employed, together with fine BZ sampling to converge the mobilities, then the Boltzmann transport equation within the RT approximation [Eq.~(\ref{eq:cond})] can yield highly accurate results over a wide temperature range.\\
\indent
%
%
%
The computed mobility increasingly deviates from the experimental result at high temperatures above 500~K. Note that for each temperature we investigate, we employ RTs computed at the same temperature, but the band velocities are obtained from DFT without accounting for finite-temperature corrections to the bandstructure.
In particular, we find that our computed mobilities are lower than experiment at $T>500$ K, a trend opposite to that shown in previous work using empirical models~\cite{Rode1970,*Rode1971}. 
A possible explanation for this difference is that most previous studies neglected the $\Gamma$$-$$L$ intervalley scattering, 
which is incorrect since the integrand in Eq.~(\ref{eq:cond}) extends well beyond $E_L$ at high temperature.\\
\indent
\begin{figure}
\centering
\includegraphics[scale=0.8]{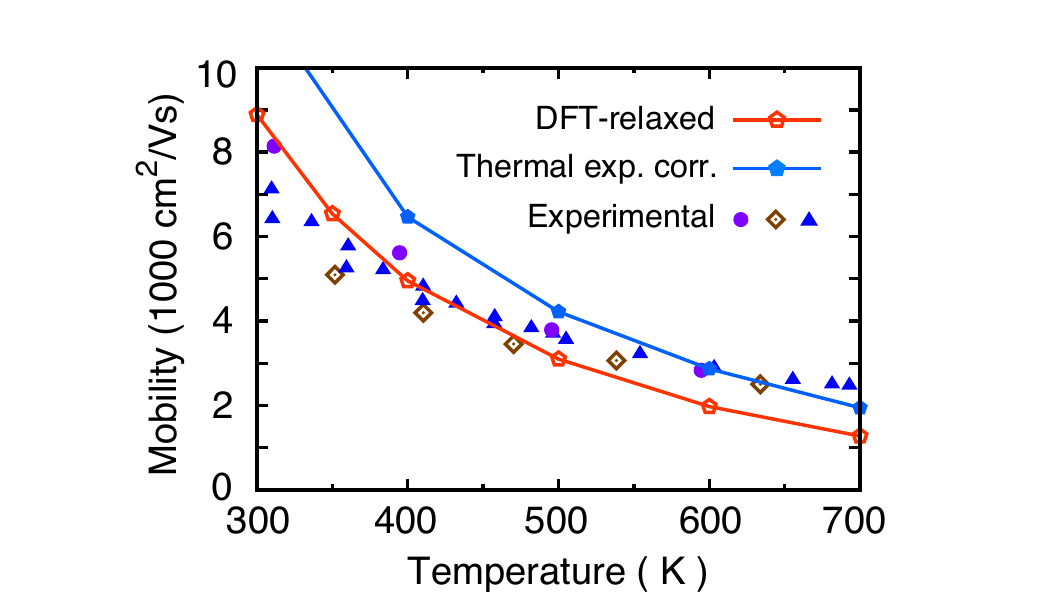}
\caption{(Color online) Electron mobilities computed using both the DFT-relaxed and thermal expansion corrected lattice parameters, for temperatures above 300~K. The sources of the experimental data are the same as in Fig.~\ref{fig4}(a). 
}\label{fig5}
\end{figure}
%
%
We attribute the deviation of our results at $T>500$ K to the lack of finite-temperature corrections to our bandstructure.  
%
To test this hypothesis, we compute the mobility using a thermal expansion corrected lattice parameter ($a \approx$ 5.57~\AA) \cite{Glazov2000}. 
Figure~\ref{fig5} shows that agreement of the computed mobility with experiment at $T>500$ K improves when we employ a lattice parameter corrected for thermal expansion. 
%
Our simple attempt to include finite temperature effects suggests that bandstructure renormalization is an important aspect of high temperature mobility calculations. 
On the other hand, we anticipate that combining ab initio temperature-dependent band structures \cite{Kawai2014,Ponce2015,Nery2016} with our accurate RT and mobility calculations would be computationally very challenging.
Lastly, we note that \textit{e}-ph matrix elements can also be derived from the $GW$ self-energy rather than from DFPT based on semilocal DFT as is done here. Recent work~\cite{Antonius2014} has shown that the \textit{e}-ph coupling strength can differ significantly in DFPT and $GW$. Future work on carrier transport should investigate this point further.\\
\indent
%
In summary, 
we demonstrate the crucial role of BZ sampling and convergence in computing the $e$-ph RTs and the mobility in polar bulk materials. The algorithms developed in this work reduce the computational cost significantly by dividing and conquering the long range part of the $e$-ph interaction, and optimizing BZ sampling.
Our mobility calculations in GaAs achieve excellent agreement with experiment, thus demonstrating that, contrary to previous results, the RT approximation of the Boltzmann equation can accurately compute the mobility in GaAs at room temperature.  
Our approach enables ab initio studies of charge transport and carrier dynamics in polar materials, with broad applications in materials science and condensed matter physics. 
The authors are working toward releasing a code to carry out the calculations shown in this work. \\
\indent
\emph{Note added.}
Recently, we became aware of a related calculation for GaAs reported by Liu \textit{et al}. \cite{Liu2016}. 
Their computed mobility within the RT approximation is significantly lower than our result, and their scattering rate in the $\Gamma$ valley is greater than ours. 
While convergence and broadening are not discussed in their work, their results are consistent with those found here for a too large Lorentzian broadening 
and for underconverged ($\sim$100$^3$) $\mathbf{k}$-point grids in the mobility calculation. These important differences lead to different conclusions in their work. \\
\newline
\indent
This work was supported by the Joint Center for Artificial Photosynthesis, a DOE Energy Innovation Hub, supported through the Office of Science of the U.S. Department of Energy under Award No. DE-SC0004993.
This research used resources of the National Energy Research
Scientific Computing Center, a DOE Office of Science User
Facility supported by the Office of Science of the U.S. Department
of Energy under Contract No. DE-AC02-05CH11231.

\bibliographystyle{apsrev4-1}
\bibliography{gaas_ref}
\end{document}